\documentclass{sf2a-conf}
\usepackage{graphicx}
\begin{document}
\TitreGlobal{SF2A2007}
\title{3D SPH simulations of grain growth in protoplanetary disks}

\author{G. Laibe}\address{Universit\'e de Lyon, Lyon, F-69003, France;
Universit\'e Lyon 1, Villeurbanne, F-69622, France;
CNRS, UMR 5574, Centre de Recherche Astrophysique de Lyon,
\'Ecole Normale Sup\'erieure de Lyon, 46 all\'ee d'Italie,
F-69364 Lyon cedex 07, France}
\author{J.-F. Gonzalez$^1$}
\author{L. Fouchet}\address{ETH Hoenggerberg Campus, Physics Department,
HPF G4.2, CH-8093 Zurich, Switzerland}
\author{S.T. Maddison}\address{Centre for Astrophysics and Supercomputing,
Swinburne University of Technology, PO Box 218, Hawthorn, VIC 3122, Australia}

\runningtitle{3D SPH simulations of grain growth in protoplanetary disks}
\setcounter{page}{1}
\index{Laibe G.}
\index{Gonzalez J.-F.}
\index{Fouchet L.}
\index{Maddison S.T.}
\maketitle
\begin{abstract}
We present the first results of the treatment of grain growth in our 3D, two-fluid (gas+dust) SPH code describing protoplanetary disks. We implement a scheme able to reproduce the variation of grain sizes caused by a variety of physical processes and test it with the analytical expression of grain growth given by Stepinski \& Valageas (1997) in simulations of a typical T Tauri disk around a one solar mass star. The results are in agreement with a turbulent growing process and validate the method. We are now able to simulate the grain growth process in a protoplanetary disk given by a more realistic physical description, currently under development. We discuss the implications of the combined effect of grain growth and dust vertical settling and radial migration on subsequent planetesimal formation.
\end{abstract} 
%
\section{Introduction}
\label{Intro}

Planets are thought to form in disks around young stars through the growth of
dust grains (Dominik et al.\ 2007). Collisions and aggregation govern the
first steps from micron-sized particles to decimetric pre-planetesimals, which
then form kilometric planetesimals via still-discussed mechanisms, and
ultimately planets. Observations of protoplanetary disks support this
mechanism by showing evidence of dust grain growth (e.g. Apai et al.\ 2004).
The understanding of this physical process is necessary to describe the first
steps of planet formation. We describe here the implementation of grain growth
in our simulations of protoplanetary disks.

\section{Physics of protoplanetary disks}
\label{SectPhysics}

Protoplanetary disks are mostly composed of gas and typically contain about
1\% of their mass in dust.
The gas evolves in a pseudo keplerian rotation due to its radial pressure
gradient, which decreases the effect of the central star's gravity. The energy
dissipation due to the gas viscosity induces an inwards mass flow and an
outwards angular momentum flow.
Dust grains only feel the gravity field of the central star, but are not
affected by either pressure gradient or viscosity and, if alone, would
therefore be in keplerian rotation.

However, gas and dust interact via a drag force which depends on their relative
velocity and slows down the dust, causing it to migrate radially inwards and to
settle vertically towards the midplane. Those processes are strongly dependent
on the size $s$ of the grains: the largest particles are totally decoupled
from the gas, the smallest ones are so strongly coupled that they follow the
gas motion, and in the intermediate regime, particles undergo settling and
migration with a varying efficiency (Barri\`ere-Fouchet et al.\ 2005).

Finally, the size of dust grains can vary: depending on their relative
velocities and material properties (Dominik \& Tielens 1997), collisions
between solid particles can make them stick and grow, or conversely break
them into smaller pieces.

\section{Grain growth in protoplanetary disks}
\label{SectSV97}

A first description of the grain growth process is given by Stepinski \&
Valageas (1997). They model a turbulent, vertically isothermal
protoplanetary disk, in which gas and dust are represented by two separate
phases interacting via aerodynamic drag in the Epstein regime. Their solid
particles are supposed to stick perfectly during collisions, and can therefore
only grow. The variation of their size $s$ is given by the following
analytical expression:
\begin{equation}
\frac{\mathrm{d}s}{\mathrm{d}t}=\sqrt{2^{^3\!/\!_2}\,\mathrm{Ro}\,\alpha}\,
\frac{\rho_\mathrm{s}}{\rho_\mathrm{d}}\,C_\mathrm{s}
\frac{\sqrt{\mathrm{Sc}-1}}{\mathrm{Sc}},
\label{EqEvol}
\end{equation}
where Ro is the Rossby number for turbulent motions, $\alpha$ the Shakura \&
Sunyaev (1973) viscosity parameter, $\rho_\mathrm{s}$ the density of matter
concentrated into solid particles, $\rho_\mathrm{d}$ the intrinsic density of
the grains, $C_\mathrm{s}$ the local gas sound speed, and Sc the Schmidt
number of the flow which estimates the effect of gas turbulence on the grains.
Sc is defined by
\begin{equation}
\mathrm{Sc}=(1+\Omega_\mathrm{k}\,t_\mathrm{s})
\sqrt{1+\frac{\bar{\mathbf{v}}^2}{V_\mathrm{t}^{2}}},
\label{EqSchmidt}
\end{equation}
where $\Omega_\mathrm{k}$ is the local keplerian velocity, $t_\mathrm{s}$
the dust stopping time, $\bar{\mathbf{v}}$ the mean relative velocity between
gas and dust, and $V_\mathrm{t}$ a turbulent velocity. The growth rate
$\displaystyle\frac{\mathrm{d}s}{\mathrm{d}t}$ depends on $s$ through the stopping time
\begin{equation}
t_\mathrm{s}=\frac{\rho_\mathrm{d}\,s}{\rho_\mathrm{g}\,C_\mathrm{s}},
\label{EqStopTime}
\end{equation}
where $\rho_\mathrm{g}$ is the gas density.

The motion of a dust grain, as mentioned in Sect.~\ref{SectPhysics}, and the
grain growth process itself presented here, which both depend on the grain
size $s$, are coupled phenomena. The understanding of the global evolution of
dust in disks therefore requires a numerical treatment.

\section{Grain growth with an SPH code}
\label{SectCode}

Our 3D, bi-fluid, Smooth Particle Hydrodynamics (SPH) code has been developed
over recent years to model vertically isothermal, non self-gravitating
protoplanetary disks. The effects of radiative transfer, magnetism,
self-gravity and turbulence are neglected. Gas and dust are treated as two
separate phases and are coupled by aerodynamic drag. The code and its results
on dust migration and settling are presented in Barri\`ere-Fouchet et al.\
(2005). It has also been applied to grain stratification in GG~Tau's
circumbinary ring (Pinte et al.\ 2007) and to gaps opened by planets in the
dust phase of protoplanetary disks (Maddison et al.\ 2007; Fouchet et al.\
2007).

The assumptions we have made are very similar to those of Stepinski \&
Valageas (1997) mentioned in Sect.~\ref{SectSV97}, their prescription for
grain growth is therefore easy to implement in our code. In contrast to our
previous work, we now allow the grain size $s$ assigned to each SPH particle,
assuming it to represent the typical size of dust grains at its position, to
vary with time following Eq.~\ref{EqEvol}. We take the initial grain size
distribution to be uniform.

\section{Results}
\label{SectRes}

\begin{figure}[t]
\begin{center}
\includegraphics{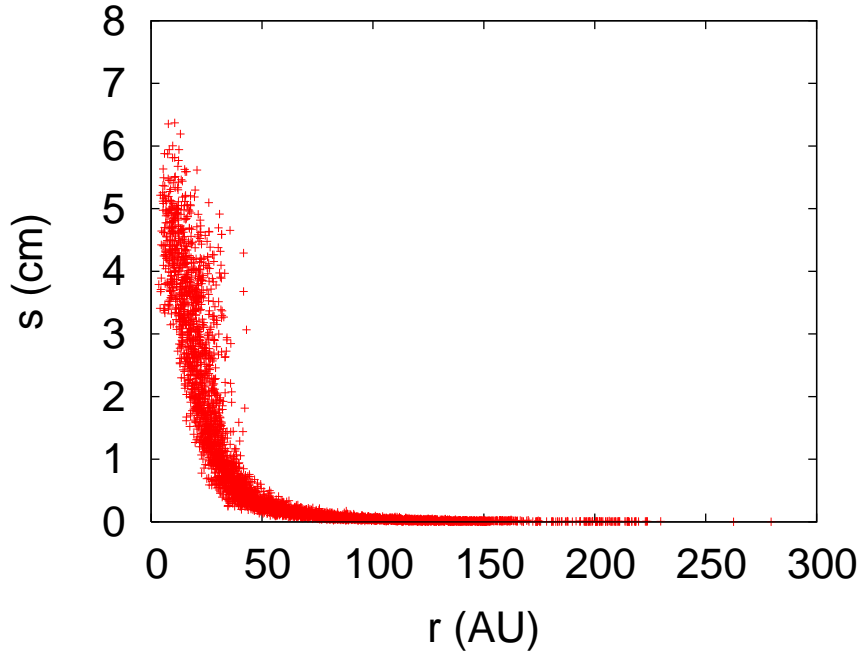}
\caption{Radial dust grain size distribution at the end of the SPH simulation
with $s_0=10\ \mu$m, in a linear scale.}
\label{FigSPHlin}
\end{center}
\end{figure}

\begin{figure}[t]
\begin{center}
\includegraphics{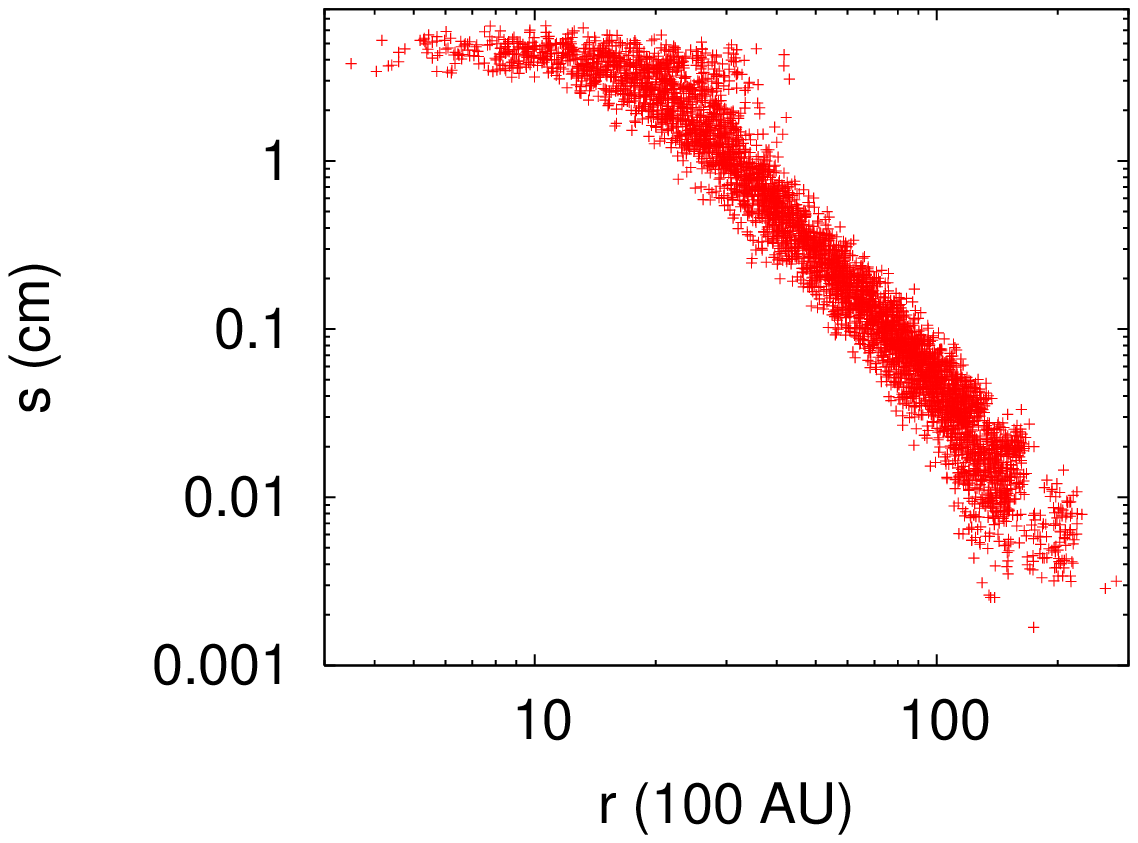}
\caption{Same as Fig.\ref{FigSPHlin} in a logarithmic scale.}
\label{FigSPHlog}
\end{center}
\end{figure}

We model grain growth in a typical T~Tauri disk of mass $M_\mathrm{disk}=0.01\
M_\odot$, with a total dust mass $M_\mathrm{dust}=0.01\ M_\mathrm{disk}$,
extending between 3 and 300~AU from the central star of mass $M_\star=1\
M_\odot$. We ran simulations with 100,000 SPH particles, evolved over 40,000
years, for a series of initial dust grain sizes ranging from $s_0=1\ \mu$m to
1~mm. Figures \ref{FigSPHlin} and \ref{FigSPHlog} show the size distribution
in the disk as a function of the radial distance to the central star at the
end of the simulation with $s_0=1\ \mu$m.

We find that grain growth occurs very quickly, especially in the inner disk
where the density is the highest (on the order of $10^{-10}$~kg.m$^{-3}$),
as one would expect from Eq.~(\ref{EqEvol}). In this region, the dust grains
reach centimetric size (see Fig.~\ref{FigSPHlin}) in only a few timesteps. In
the outer parts of the disk, where the density is far lower (as low as
$10^{-15}$~kg.m$^{-3}$), the grains grow much more slowly and their size stays
below the millimeter (see Fig.~\ref{FigSPHlog}).

At any place in the disk, the growth timescale is far smaller than the
migration timescale. Hence, after a transitory regime where grains grow from
their uniform initial size, a quasi-stationary regime quickly takes place.
Indeed, an inwards migrating grain coming into a denser region grows almost
instantaneously to a size characteristic of the grains at its new position,
which in turn evolves slowly over time. The global shape of the plots shown
in Figs.~\ref{FigSPHlin} and \ref{FigSPHlog} stays the same as time goes on,
but the size distribution slowly progresses to larger sizes.

The initial size $s_0$ has very little influence on the final size
distribution, and in particular on the maximum size reached. Indeed, for a
given dust disk mass, the smaller the dust grains, the more numerous they are,
and the more frequent their collisions are. Consequently, even if the initial
grains are very small, they will very quicky reach the asymptotic regime
and lead to centrimetric grains in the central parts.

Our results are consistent with those obtained by Dullemond \& Dominik (2005).
With a model solving the coagulation equation in presence of turbulence, they
found a very fast grain growth in T~Tauri protoplanetary disks, depleting very
small sizes and producing centimetric grains.

\section{Tests with a semi-analytical model}
\label{SectTest}

In the Stepinski \& Valageas (1997) prescription, presented in
Eqs.~(\ref{EqEvol}) and (\ref{EqSchmidt}), the growth rate depends on
the local differential velocity between gas and dust. This quantity is
determined by the SPH code. In order to test the validity of our
implementation, we develop a semi-analytical model of viscous disks, to which
we compare our SPH simulations. In this model, as in our code, the disk is
assumed to be vertically isothermal, non turbulent, non self-graviting and all
magnetic effects are neglected. To describe the gas migration, the viscous
character of the gas has to be considered. For simplicity, the gas flow is
incompressible and in a stationary regime, and the migration timescale is
assumed to be large compared to all other timescales. Under these hypotheses,
the gas radial velocity is proportional to $-\displaystyle\frac{\nu}{r}$,
where $\nu$ is the gas kinematic viscosity.

The equation of dynamics for the grains is:
\begin{equation}
\frac{\mathrm{d}\mathbf{v}_\mathrm{d}}{\mathrm{d}t}=
-\frac{\mathbf{v}_\mathrm{d}-\mathbf{v}_\mathrm{g}}{t_\mathrm{s}}+\mathbf{g},
\label{EqDustDyn}
\end{equation}
where the d and g subscripts refer to the dust grains and the gas,
respectively. For small grains, we derive an expression for the components of
the differential velocity (Laibe \& Gonzalez 2008). Taking into account the
initial conditions for the dust, which is injected with the same local
velocity than the gas, i.e.\ $\mathbf{v}_{d}(t=0)=\mathbf{v}_{g}(t=0)$,
we obtain
\begin{equation}
\left\{\begin{array}{rcl}
v_{\mathrm{d},r}-v_{\mathrm{g},r} & = & -t_\mathrm{s}\displaystyle\left(\frac{v_{\mathrm{g},r}^{2}}{r}-\frac{1}{\rho}\frac{\mathrm{d}P}{\mathrm{d}r}\right)\left(1-e^{-t/t_\mathrm{s}}\right)\\
v_{\mathrm{d},\theta}-v_{\mathrm{g},\theta} & = & 0\\ \\
v_{\mathrm{d},z}-v_{\mathrm{g},z} & = & 0
\end{array}\right. .
\label{EqDeltaV}
\end{equation}
Using the $\displaystyle\frac{\nu}{r}$ dependence of the gas radial velocity,
we have rewritten the viscosity term so that it appears only implicitly in the
$\displaystyle\frac{v_{\mathrm{g},r}^{2}}{r}$ term. Therefore, both the
viscosity and the radial pressure gradient contribute to the radial
differential velocity.

\begin{figure}[t]
\begin{center}
\includegraphics{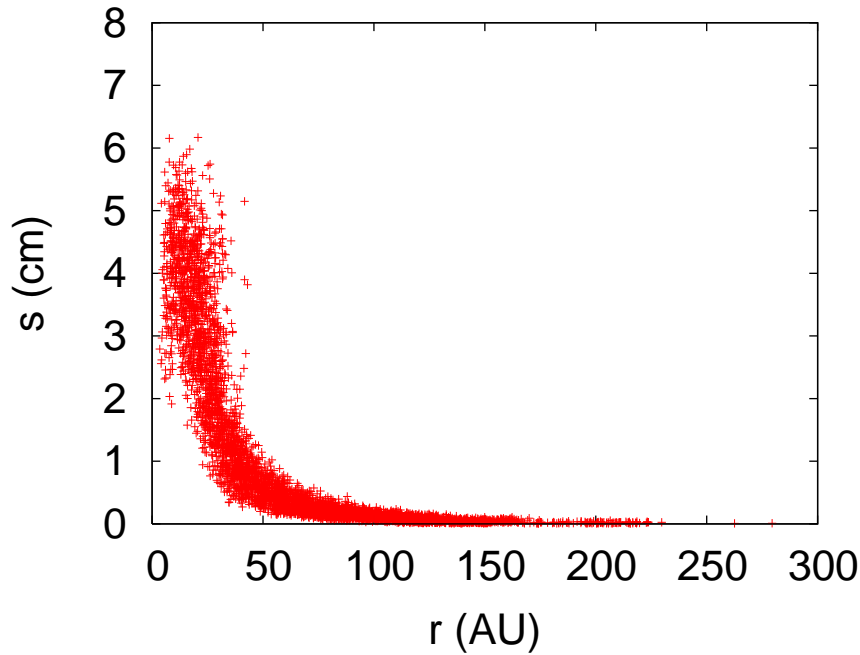}
\caption{Radial dust grain size distribution obtained by the analytical model.}
\label{FigTestComplet}
\end{center}
\end{figure}

We ran a new simulation with the same settings as in Sect.~\ref{SectRes},
but this time $\bar{\mathbf{v}}$ in Eq.~(\ref{EqSchmidt}) is computed
via Eq.~(\ref{EqDeltaV}) instead of using the individual gas and dust
velocities. Figure~\ref{FigTestComplet} displays the resulting grain size
distribution, it resembles the one shown in Fig.~\ref{FigSPHlin}, in
particular the maximum grain size reached at the inner disk edge is of the
same order of magnitude, about 6~cm. This provides a consistency check of our
implementation in the SPH code. The residual difference between the two
simulations is due to the restricting hypotheses made in the analytical
model.

\begin{figure}[t]
\begin{center}
\includegraphics{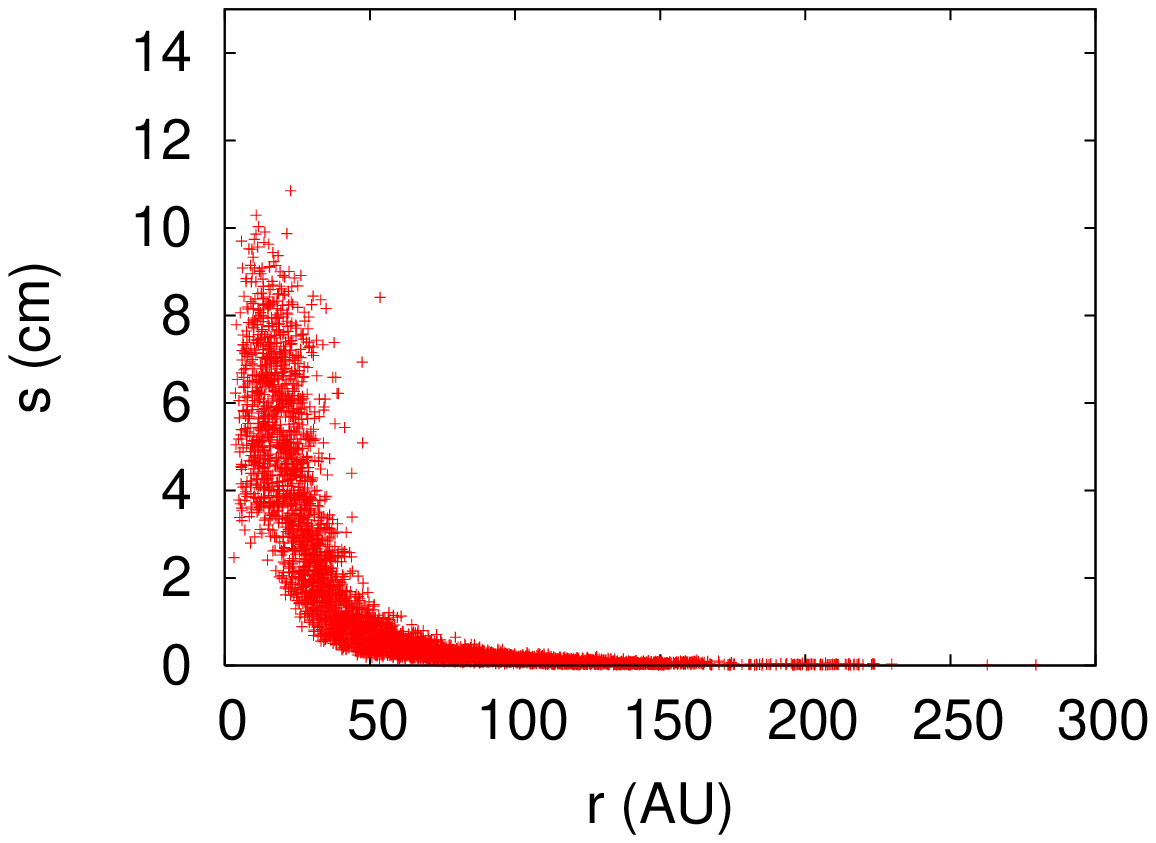}
\caption{Radial dust grain size distribution obtained by the analytical model
when the viscosity term is neglected.}
\label{FigTestSansVisc} 
\end{center}
\end{figure}

In order to investigate the relative importance of the two terms contributing
to the radial differential velocity, we ran a simulation omitting the viscosity
term. As illustrated in Fig.~\ref{FigTestSansVisc}, the maximum grain size in
the inner disk is about twice as large as the one obtained in both previous
simulations. This shows the importance of the viscosity term in
Eq.~(\ref{EqDeltaV}), and that it should not be neglected.

\section{Conclusion}
\label{Concl}

The physics of grain growth in protoplanetary disks is complex and involves
coupled processes, it therefore requires the use of numerical simulations.
We have implemented in our 3D SPH code a mechanism able to treat this grain
growth and validated it through the use of the simple model of Stepinski \&
Valageas (1997) and comparisons with a semi-analytical model we developed.

In accordance with physical intuition, dust grain grow much more quickly in the
denser, central regions of the disk, where centrimetric sized are reached.
The growing time is smaller than the migration time and a quasi-stationnary
distribution of grain size appears in the disk. The small grains are depleted
too rapidly to be consistent with observations of protoplanetary disks,
showing the need to take into account other processes such as shattering of
large grains in high-velocity collisions.

In order to treat a more realistic grain growth process, we are working on a
more detailed model taking into account microscopic interactions
between the grains, kinetic energy dissipation, porosity, and re-fragmentation,
from which we will derive a prescription to include in our code in the same
way as the one we have presented here. We also plan to add a simple, but
realistic, treatment of turbulence.

\end{document}